\documentclass[preprint,12pt]{elsarticle}

\usepackage{amssymb}
\usepackage[labelfont=bf]{caption}
\usepackage{subcaption}
\usepackage{graphicx}
\usepackage{booktabs}
\usepackage{color,colortbl}
\definecolor{Gray}{gray}{0.9}
\journal{Journal of Computational Science}

\begin{document}

\begin{frontmatter}


\title{Non-local Potts model on random lattice and chromatic number of a plane}
\author[1]{V. Shevchenko}
\author[2]{A. Tanashkin\corref{cor1}}
	\ead{tanashkin.as@dvfu.ru}
\cortext[cor1]{Corresponding author}
\affiliation[1]{organization={National Research Nuclear University MEPhI},
		 addressline={Kashirskoe sch. 31},
		 city={Moscow},
		 postcode={115409},
%		 state={},
		 country={Russia}}

\affiliation[2]{organization={Pacific Quantum Center},
		 addressline={FEFU Campus, 10 Ajax Bay, Russky Island},
		 city={Vladivostok},
		 postcode={690922},
%		 state={Prymorie},
		 country={Russia}}

%%%%%%%%%%%%%%%%%%%%%%%%%%%%%%%%%%%%%%%%%%%%%%%%%%%%%%%%%%%%%%%%%%%%%%%%%
%\title{Non-local Potts model on random lattice and chromatic number of a plane}
%
%%% use optional labels to link authors explicitly to addresses:
%\author[label1]{V. Shevchenko}
%\author[label2]{A. Tanashkin}
%	\ead{tanashkin.as@dvfu.ru}
%
%\affiliation[label1]{organization={National Research Nuclear University MEPhI},
%		 addressline={Kashirskoe sch. 31},
%		 city={Moscow},
%		 postcode={115409},
%%		 state={},
%		 country={Russia}}
%
%\affiliation[label2]{organization={Pacific Quantum Center},
%		 addressline={FEFU Campus, 10 Ajax Bay, Russky Island},
%		 city={Vladivostok},
%		 postcode={690922},
%%		 state={Prymorie},
%		 country={Russia}}
%%%%%%%%%%%%%%%%%%%%%%%%%%%%%%%%%%%%%%%%%%%%%%%%%%%%%%%%%%%%%%%%%%%%%%%%%%5
%\author{}

%\affiliation{organization={},%Department and Organization

\begin{abstract}

Statistical models are widely used for investigation of complex system's behavior.
Most of the models considered in the literature are formulated on regular lattices with nearest neighbor interactions. The models with non-local interactions have been less studied. We investigate in the present article an example of such a model -- non-local $q$-color Potts model on a random $d=2$ lattice, where only the same color spins at unit distance (within some small margin $\delta$) interact.
We analyze numerically the structure of the vacuum states in this model and discuss qualitative features of the corresponding patterns. Conjectured relation with the chromatic number of a plane problem is discussed.
\end{abstract}




\end{frontmatter}


%\section{}
%\label{}

\section{Introduction}
\label{intro}

Statistical models on discrete lattices are proved to be an extremely useful tool in studying various problems in fundamental and applied science.
The best known is certainly the Ising model and its generalizations like the Potts models family \cite{Potts-review,Potts-adv}.
As is well known, the typical model of this kind is defined by the functional
\begin{eqnarray}
H = \sum_{i, k \in M}\> J_{ik} \> \sigma_i \cdot \sigma_k,
\label{potts-e1}
\end{eqnarray}
where the dynamical variables $\sigma_i$ - "spins" - take values in some set $S$, while the indices $i$, $k$, run over the lattice $M$.
The matrix $J_{ik}$ encodes the "interaction strength" between sites $i$ and $k$.
It is usually assumed that $S, M, J_{ik}$ and the composition law are such that $H$ is a real non-negative number for any spin configuration.

Being originally invented to study magnetic phenomena, the area of applications of the models (\ref{potts-e1}) has become remarkably broad. Of special interest are links with complex behavior of many-body systems, ranging from studies of collective decision making \cite{PhysRevLett.123.068101} and computer vision \cite{Nonlocal-potts} to biological evolution (see \cite{2001} and references therein) and airline crew scheduling \cite{article-A}. In all these cases the phase structure of the models and their phase transition dynamics dependence on effective temperature or internal parameters (see \cite{article-B} and references therein) is of interest as well as the minimum energy states and their structure.

For many cases it is convenient to consider the Potts-type models in the mean field approximation (see e.g. \cite{Sportiello}), where standard machinery of statistical field theory, including path integral representation \cite{Kholodenko} is well suited. However, generally, speaking minimizing the Potts model energy is known to be NP-hard problem and cannot be solved exactly in reasonable time.
Therefore one uses various approximations and numerical methods to approximate the exact
solution as effectively as possible.

%The two most common problems are finding the vacuum state of (\ref{potts-e1}), i.e. the configuration(s) %$\{ \sigma_i \}$, which provide(s) minimum of $H$, and computation of the corresponding partition function
%\begin{eqnarray}
%Z = \sum_{\{ \sigma_i \}}\> e^{-\beta H}.
%\label{kk}
%\end{eqnarray}
%In the low temperature limit the latter one is dominated by the vacuum state(s) contribution, but needless %to say that computing (\ref{kk}) and studying its $\beta \to \infty$ limit is usually by far not the most %economic way to find the vacuum state(s) of a model.

The most peculiar feature of all these models is a subtle interplay between a structure of the space $S$ the degrees of freedom belong to, and geometry and topology of the bulk, encoded in $M$ and the interaction kernel $J_{ik}$.
In particular, if the model in question undergoes phase transition at some value of the inverse temperature $\beta$, the typical correlation length in the vicinity of this point can be much larger than the scale of the microscopic structure of the lattice.
As a result, large distance correlators become insensitive to this structure and can correspond to some continuous theory in this limit, which shares with the original model only some global parameters like dimension or topology.
However if one is far from the phase transition point, the lattice microstructure is manifest.
For example, in the ground state of the standard nearest neighbor antiferromagnetic Ising model the typical correlation length equals to just one lattice link and the concrete pattern of the spins in this lowest state depends on the geometry of the lattice (hypercubic, honeycomb etc).

Instead, we consider random lattice in this paper, which is another well known way to get rid of the dependence on the lattice microstructure, besides considering the model in the continuum \cite{article-E}. The price to pay is an absence of a fixed geometry.
Therefore the only geometric invariants can be encoded in the coupling pattern.
We consider two-dimensional case with the interaction kernel, describing interaction of the given spin with all spins, whose distance from it is larger than $R-\delta/2$ but smaller that $R+\delta/2$:
\begin{eqnarray}
J_{ik} = \left\{ \begin{array}{ll}  J, & \; \; R-\frac{\delta}{2} \le |i-k| \le R+\frac{\delta}{2} \\
0, &\;\; \mbox{otherwise.} \end{array}  \right.
\label{ker}
\end{eqnarray}
We will assume that $\delta \ll R$, so the interaction region can be thought of as a ring.

For a regular lattice the above would mean that the number of neighbors is the same for any spin.
For example, in the $d$-dimensional Ising model on a hypercubic lattice with the link $R$ it is exactly $2d$ for any $0<\delta \ll R$.

For the random lattice, let us take $N$ sites to be randomly distributed over a compact $d$-dimensional region with linear size $L$, so we get $\rho = N/L^d$ for the average site density.
Then the average number of each spin neighbors is given by $\rho\sim R^{d-1} \delta $.
It is easy to see that if one requires that each spin to have at least one neighbor on average, there will be $\sim R^d \rho \gg 1$ non-interacting spins, which are closer to the chosen one, than the neighbors it interacts with. This is the most peculiar difference between  statistical models on regular and random lattices: interaction at the fixed (minimal) distance (within perhaps some margin $\delta$) is the same thing as interaction with the nearest neighbors for the former, but not for the latter ones.

We proceed below with the analysis of (\ref{potts-e1}) on a plane ($d=2$) with the
interaction kernel (\ref{ker}).
The spins $\sigma_i$ are represented by $q$-component vectors (we use the word "colors" for different components of these vectors), $\sigma_i = (\sigma_i^1, \sigma_i^2, ..., \sigma_i^q)$ where $\sigma_i^k \in \{0,1\}$ and $\sum_{k=1}^q  \sigma_i^k = 1$, so the product
$
\sigma_i \cdot \sigma_k = 1 $ if spins in the sites $i$ and $k$ have the same color, and $
\sigma_i \cdot \sigma_k = 0 $ otherwise.

The idea of nearest neighbor interaction goes to physical interpretation of the degrees of freedom as spins whose interaction decreases with the distance and hence is the strongest between the closest spins.
The non-local interaction models have attracted much less attention in the literature.
The motivations for considering such models come from biological, social and technical networks that are neither completely random nor completely regular. One known example is the small-world topology \cite{WS} which is believed to describe some properties of the brain network, besides others. Three-state Potts model on non-local directed small-world lattices is studied in \cite{Article-D}. Closely related are models with finite size degrees of freedom. In our context let us mention studies of two-dimensional crystals from 5-fold-symmetric molecules \cite{2Dcrystals}, Monte Carlo studies of hard pentagons \cite{MC-pentagons}, studies of systems of rotationally asymmetric hard kite particles \cite{Kite-particles}. In these and other studies interesting patterns have been observed, which are not reproducible by local models. While in these papers non-locality had its roots in the physical properties of the corresponding systems, it was used as an artificial regulator in \cite{Nonlocal-potts}. It was shown that non-local interaction kernel smoothes the boundaries between spatial regions having different (color) Potts labels, which is advantageous in some computer vision applications.

\begin{figure}[t]
	\centering
	\includegraphics[width=1\linewidth]{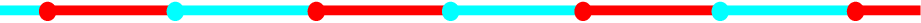}
	\caption{{\small{The solution of EHN problem for one dimension -- the alternating lines of unit length with one excluded point (at the right edge in our example).}}}
	\label{fig:1Dcont}
\end{figure}

\begin{figure}
     \centering
     \begin{subfigure}[b]{0.15\textwidth}
         \centering
         \includegraphics[width=\textwidth]{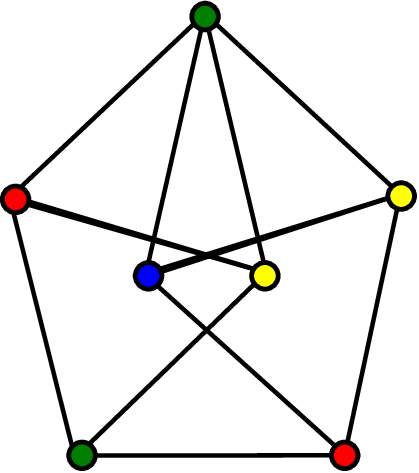}
         \caption{}
			 \label{fig:moser}
     \end{subfigure}
	 \hspace{20mm}
     \begin{subfigure}[b]{0.2\textwidth}
         \centering
         \includegraphics[width=\textwidth]{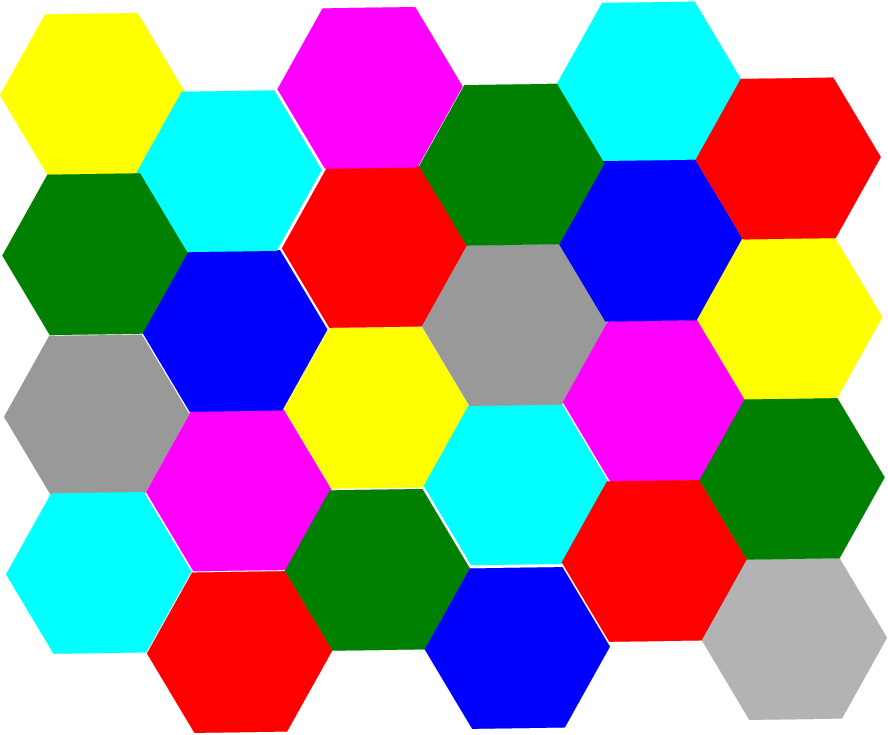}
         \caption{}
         \label{fig:7colors}
     \end{subfigure}
		\caption{{\small{Illustration of the fact that $4\leqslant q \leqslant 7$: \textbf{(a)} The Moser's spindle -- four-chromatic unit distance graph on 11 vertices proving the necessity of minimum 4 colors for coloring plane such that there are no points of the same color at the unit distance apart; \textbf{(b)} The coloring of the plane with 7 colors such that there are no points of the same color at the unit distance apart.}}}
        \label{fig:estimators}
\end{figure}

In the present paper we consider the model with interaction at {\it finite} distance, i.e. the spins interact when they are located within some finite distance from each other.
This model can be seen as a discretized version of the combinatorial topology problem of unit graph coloring. This problem is known in mathematical literature under the name Hadwiger-Nelson problem (hereafter referred to as EHN problem, E stands for Paul Erdos who made significant contribution to this topic \cite{Erdos1,Erdos2}) and can be formulated as follows: what is the minimal number of colors $q$ one has to use to color the $\mathbf{R}^d$ space in such a way that no two points at unit distance are colored identically \cite{Hadwiger}.
There is a vast mathematical literature on the subject \cite{Erdos1,Erdos2,MathBohemia,Moser}.
Let us summarize the main findings here. In the $d=1$ case the EHN problem has a trivial solution  $q=2$ and the minimal configuration is shown in the figure~\ref{fig:1Dcont}.
Surprisingly, already for a plane ($d=2$) an answer is unknown.
It can be easily shown that $q=3$ colors is not enough (figure~\ref{fig:moser}), while for $q=7$ a regular partition satisfying the requirement exists (figure~\ref{fig:7colors}).
Of course, the solution shown in the figure~\ref{fig:7colors} is not unique.
Quite recently it was argued with the help of a sophisticated mathematical construction that $q=4$ is also too small \cite{Grey}.
So, one is left with three possibilities: $q=5$, $q=6$ or $q=7$.

The intrinsic mathematical difficulty of the EHN problem has its roots in the necessity to assign a discrete index of color $k=1,..,q$ to each point of the continuum set $\mathbf{R}^d$.
The minimal configuration for $q=7$ is characterized by clustering of the plane in finite size domains, which renders the problem from continuum to countable set.
It could happen that other kinds of solutions do exist (for the same value $q=7$ or for smaller $q$), where the points on the line connecting any two points (however close) have colors, different from the colors of the edge points. Such "fractal colorings" have no intrinsic ultraviolet scale.

In the settings (\ref{potts-e1}), (\ref{ker}) a limit corresponding to the original EHN problem would be recovered as $N\to \infty$, $L\to \infty$, $\delta \to 0$, while $R=1$ without loss of generality.
The EHN problem in this language is to find (for a given $q$) a configuration corresponding to $H=0$.

\section{The method}
\label{sec:1}
\subsection{Setting up the problem}
\label{subsec:1}

\begin{figure}[t]
	\centering
	\begin{subfigure}[p]{0.45\textwidth}
		\centering			
		\includegraphics[width=\textwidth]{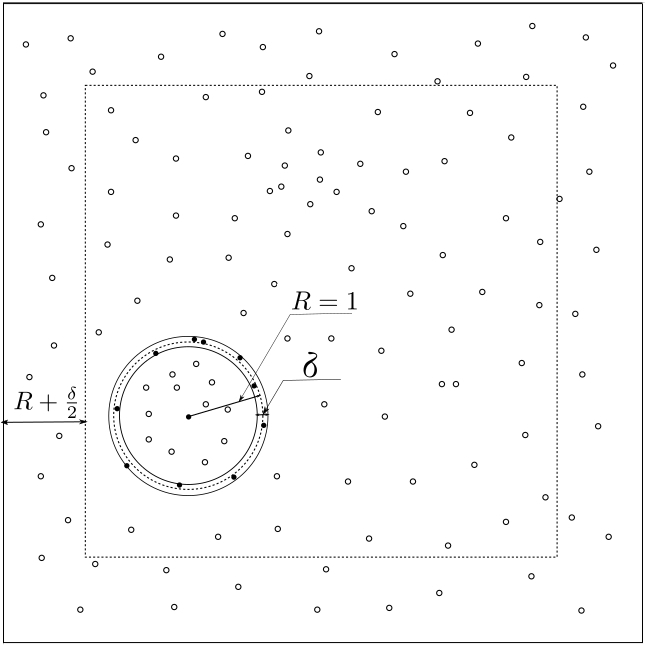}
		\caption{}
		\label{fig:model}
	\end{subfigure}
	\hfill
	\begin{subfigure}[p]{0.45\textwidth}
		\centering
		\includegraphics[width=\textwidth]{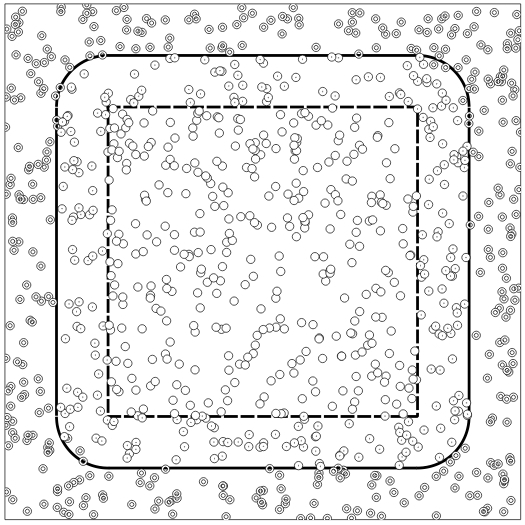}
		\caption{}
		\label{fig:energy}
	\end{subfigure}
		\caption{{\small{\textbf{(a)}: The schematic picture of the model. A particle at the center of the ring interacts only with its ring-neighbors, which are, by definition, particles at $R \pm \delta/2$ distance from the chosen one. The dotted square represents the area outside of which the colors of particles are fixed - the implementation of fixed boundary conditions; \textbf{(b)}: The schematic picture of energy zones. All particles have the same properties but are depicted differently depending on the zone they belong to. The inside energy is calculated only between open circled particles and each pair of them counted twice. Pairs which consist of one open circled particle from inside zone and dotted circled particle from adjoining zone counted once. They contribute to outside energy. The interaction of pairs of dotted circled particles doesn't contribute to energy. Interactions with ring symbol particles also omitted. It is worth mentioning that one should not be deluded by sparse density of particles - in reality the density is much higher and the example is given only for illustrative purposes.}}}
		\label{fig:schemes}
\end{figure}

First, let's make some refinements in the equation (\ref{potts-e1}) describing the energy function $H$ of the system. It can be normalized by introducing the normalization factor $A$:
\begin{eqnarray}
E = A \sum\limits_{x,y} J_{xy} \delta_{i(x)i(y)}
\label{E-red}
\end{eqnarray}
and we assume $A = 1$ in what follows.
The alternative choice $A = \frac{L^2 q}{2\pi R \delta N^2}$ would correspond to the energy of random configuration of the order of unity.
In the expression (\ref{E-red}) interaction kernel is given by (\ref{ker}), $i(x)$ -- the color of the site $x$, $\delta_{i(x)i(y)}$ -- Kronecker delta.
The schematic picture is shown in the figure \ref{fig:schemes}.
The important parameter of this model is the averaged number of interacted neighbors for each particle. It is given by relation
\begin{eqnarray}
	\langle n \rangle = 2\pi\delta\frac{N}{L^2}.
	\label{nneigh}
\end{eqnarray}

In our work we fix this number as $\langle n \rangle = 50$.
This number is large enough to observe the basic properties of the model but at the same time allows to perform computations in reasonable time.
Also we fix the linear size of area $L = 20$ and width of interaction ring $\delta = 0.02$.
This corresponds to the number of sites equal to $N = 159\,155$.
All our simulations are based on this setup.
%===========improvement=============
The choice of parameters is discussed in the section \ref{paramchoice}.
%===========improvement=============

The true minimum $E=0$ is reached only if there are no particles of the same color that are ring-neighbors and, as should be clear from the above discussion, there is no way to get $E=0$ for small values of $q$.
There are two questions in this respect: first, what are the typical patterns for minimized $E$ (for small as well as for large $q$) and, second, for which $q$ the energy $E$ approaches zero (perhaps, within statistical errors).

Another important thing is the boundary conditions.
The most common are periodic conditions.
However in our case it is not suitable because of the non-local nature of the model.
The non-locality $R$ leads to artificial effects depending on whether $L/R$ is integer or not, where $L$ is the linear size of the domain.
Therefore fixed boundary conditions, illustrated in the figure \ref{fig:model}, are more convenient.
Particles outside the dotted area stay unchanged through the process of minimization.
Due to the fact that this belt of particles with fixed color makes an influence on particles inside, the energy of the system should be calculated at some distance from the border.
We have chosen to compute energy in the internal region with the size 11 $\times$ 11 (dotted line in the figure \ref{fig:energy}).
The energy $E$ is the sum of two components: the energy $E_{\mathrm{ins}}$ of interaction only between particles that are inside energy area and the energy $E_{\mathrm{out}}$ of interaction of particles located inside area with particles outside.
The schematic picture is represented in the figure \ref{fig:energy}. The total energy $E$ is given by

\begin{equation}
	E = 2 E_{\mathrm{ins}} + E_{\mathrm{out}}.
\label{energy}
\end{equation}

It should be pointed out that energies $E_\mathrm{ins}$ and $E_\mathrm{out}$ are both non-negative quantities.
Therefore vanishing of total energy in the given region implies vanishing of energy in any internal area of it.

\subsection{The algorithm}
\label{subsec:2}
The simplest possible algorithm is the greedy one, which accepts the new configuration only if its energy is smaller or equal.
The well known drawback of this procedure is the tendency of the system to fall into the closest state, corresponding to the local minimum, which can be quite far from the global minimum.
The simulated annealing algorithm \cite{sa} suits much better for this task and differs from the greedy algorithm by a possibility of accepting a state with higher energy.
The probability of such an event decreases over time and eventually system comes to the state which we consider as a vacuum state.
The algorithm starts from the initial configuration of randomly distributed and colored particles and goes through all of them in random order.
The positions of particles are fixed while their colors are changed to random color at every step.
If the energy decreases or stays unchanged then a new color is accepted.
But if the energy increases then new color is accepted with probability $P = \exp[(E - E^\prime)/T]$, where $E$ is the energy of current state, $E^\prime$ is the energy of new state and $T$ is artificial temperature which decreases as some function of the step number. By trying different decrease functions we have chosen linear dependence, corresponding to $T_{n+1} = T_n - \Delta T$.
After visiting all particles the $T$ is decreased and the above process is repeated until the temperature reaches zero. As the algorithm is running the probability of accepting configuration with higher energy is going down making harder for the system to leap between minima.

%===============improvement====================
\begin{figure}
	\centering
		\begin{subfigure}[b]{0.49\textwidth}
			\centering
			\includegraphics[width=\textwidth]{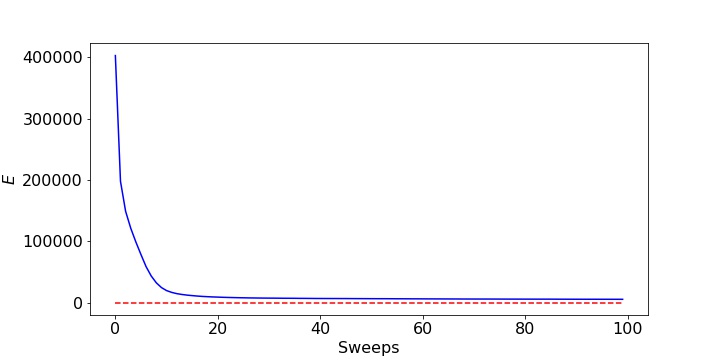}
			\caption{}
				\label{fig:evo_greedy}
		\end{subfigure}
		\hfill
		\begin{subfigure}[b]{0.49\textwidth}
			\centering
			\includegraphics[width=\textwidth]{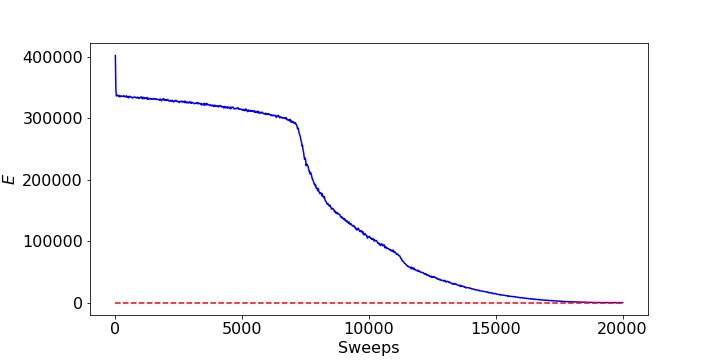}
			\caption{}
			\label{fig:evo_anneal}
		\end{subfigure}
		\caption{{\small{Typical energy minimization curves for $q=6$. \textbf{(a)}: Greedy algorithm; \textbf{(b)}: Simulated annealing}}}
	\label{fig:evo}
\end{figure}

The fundamental difference between greedy and simulated annealing algorithms is shown in the figure \ref{fig:evo} where energy evolution curves are depicted.
At the left energy decreases extremely over only a few dozens of sweeps (by sweep we mean going throw all particles once) and the system ends up in the local minimum state from which energy cannot be reduced sufficiently -- all subsequent algorithm's iterations just slightly refine the value of energy.
By comparison, simulated annealing minimization represented at the right is a fairly long (in terms of required sweeps) process having some stages.
The total number of efficient steps is by several orders of magnitude higher than for greedy minimization.
First, its behavior is similar to greedy algorithm but then very quickly changes to slow decreasing constituting about one third of the whole minimization procedure.
Next, the phase transition happens around the point of 7500 steps.
At this moment clusters of particles start to emerge and the energy again decreases very quickly.
The analogy with the process of crystallization can easily be drawn by referring to corresponding artificial temperature as "crystallization temperature".
The point of phase transition shifts depending on the value of $q$.
Finally, there are some fluctuations at the borders of formed clusters following the further decrease of energy.
As a result, the latter algorithm leads to much lower energy than the former one.
%===============improvement=====================
\section{Results }
\label{sec:2}

\begin{table}
	\centering
	\caption{The parameters of simulated model}
	\begin{tabular}{@{}lll}
		\toprule
		%\rowcolor{Gray}
		Quantity & Description & Value\\
		\midrule
		$R$ & radius of interaction & 1.0\\
		%\hline
		$\delta$ & width of interaction & 0.02\\
		%\hline
		$L$ & linear size of area & 20.0\\
		%\hline
		$N$ & the number of particles & $159\,155$\\
		%\hline
		$q$ & the number of colors & 2..7 \\
		%\hline
		$l_\mathrm{f}$ & the width of fixed boundary & 1.01\\
		%\hline
		$\delta / R$ & localization of interaction &   0.02\\
		%\hline
		$L/(R\sqrt{N})$ & ratio of average distance
		between points to $R$ & 0.05\\
		%\hline
		$2\pi R N \delta / L^2$ & average number of neighbors
		the spin interacts with &  50\\
		\toprule
	\end{tabular}
	\label{tab:1}       % Give a unique label
\end{table}
Relevant parameters of our model are summarized in the table \ref{tab:1}.
For each choice of $q$ we made 200 runs from random configurations and every run generated new random coloring of particles keeping positions the same.
We also checked the case when positions were generated randomly for each run but did not find any difference.
Below we present the results for different number of colors.

\subsection{$q\leqslant4$}
\begin{figure}
	\centering
		\begin{subfigure}[b]{0.45\textwidth}
			\centering
			\includegraphics[width=\textwidth]{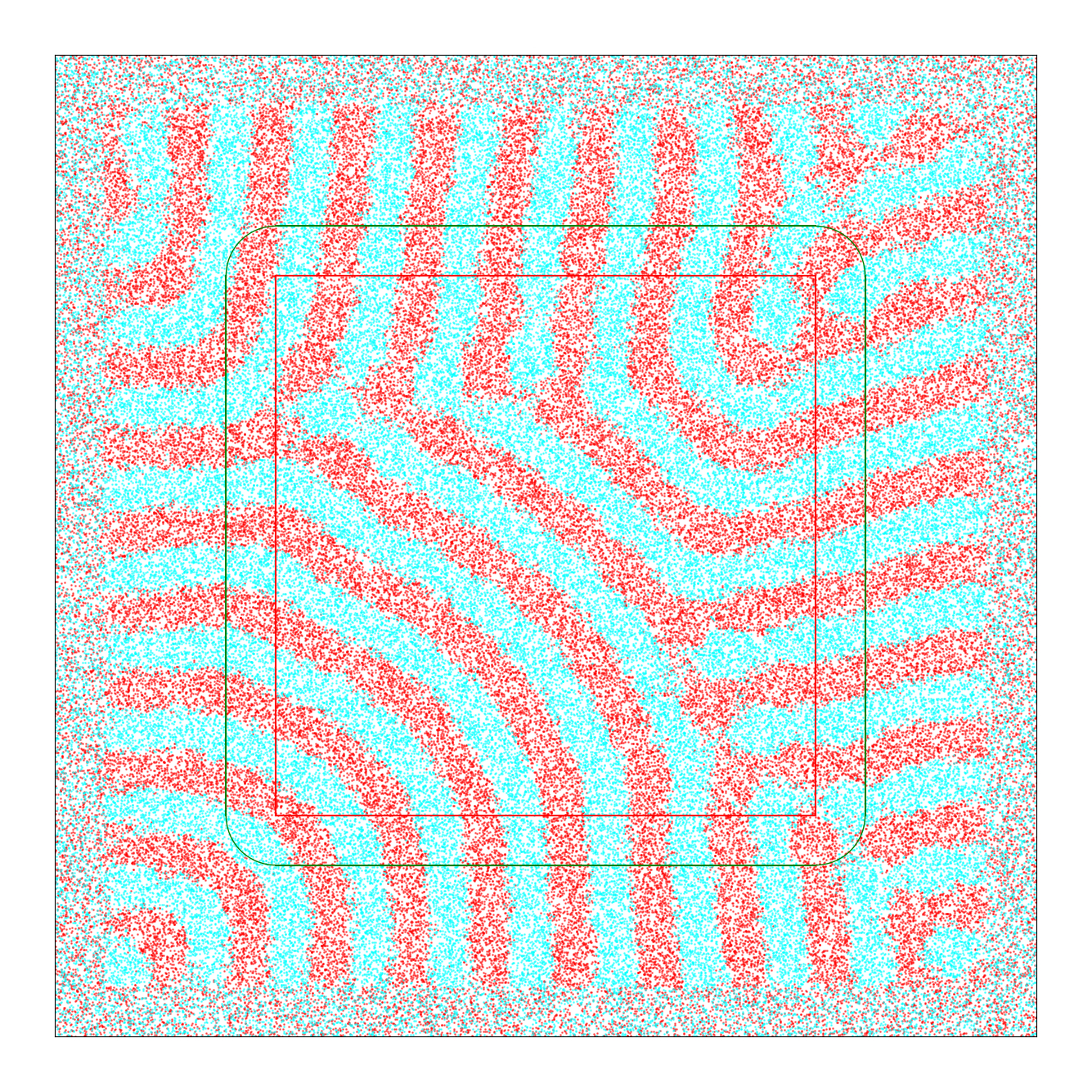}
			\caption{}
				\label{fig:q2}
		\end{subfigure}
		\hfill
		\begin{subfigure}[b]{0.45\textwidth}
			\centering
			\includegraphics[width=\textwidth]{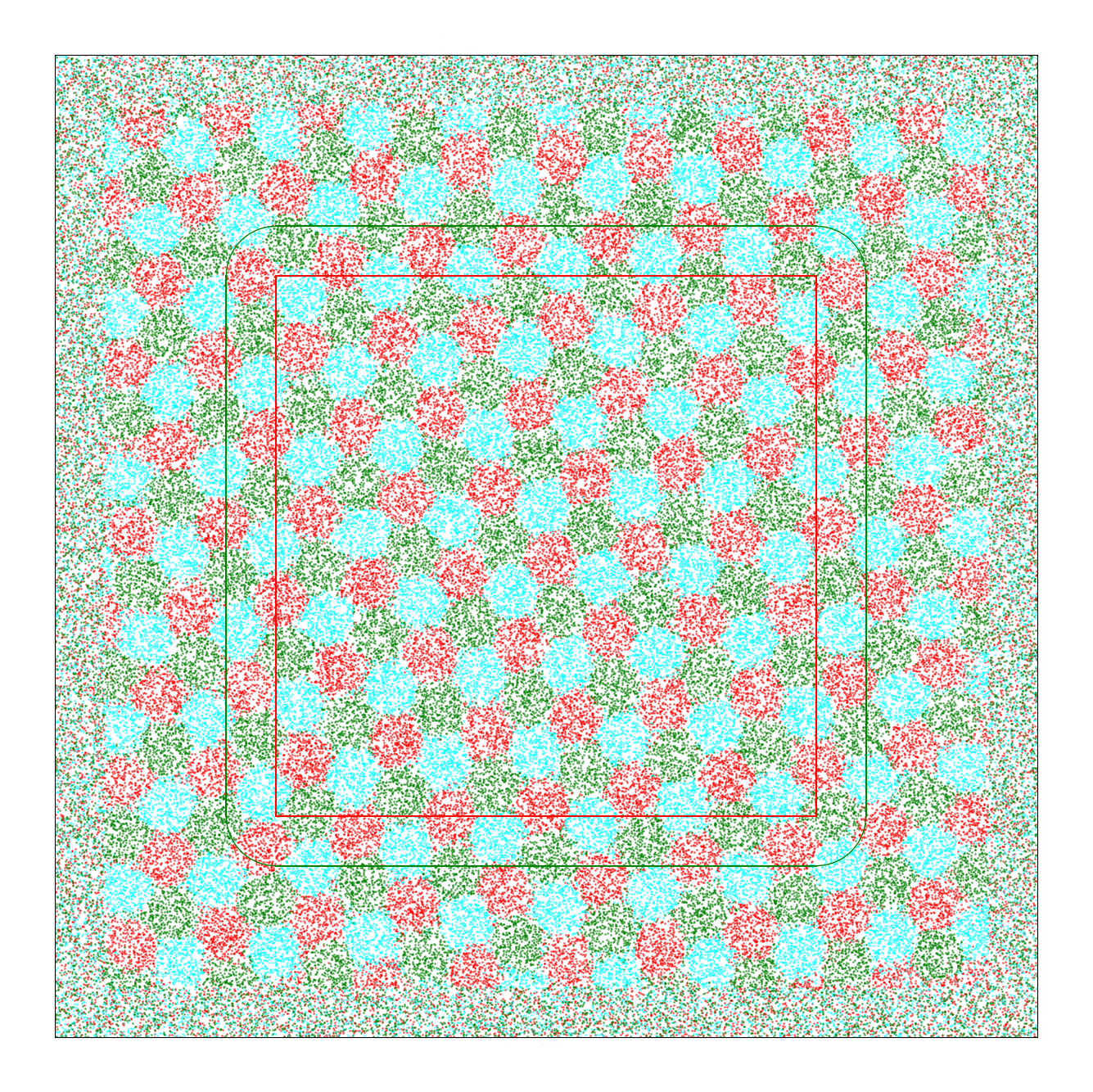}
			\caption{}
			\label{fig:q3}
		\end{subfigure}
		\caption{{\small{Typical vacuum configurations. \textbf{(a)}: $q$ = 2; \textbf{(b)}: $q$ = 3}}}
	\label{q23}
\end{figure}

\begin{figure}
	\centering
		\begin{subfigure}[b]{0.45\textwidth}
			\centering
			\includegraphics[width=\textwidth]{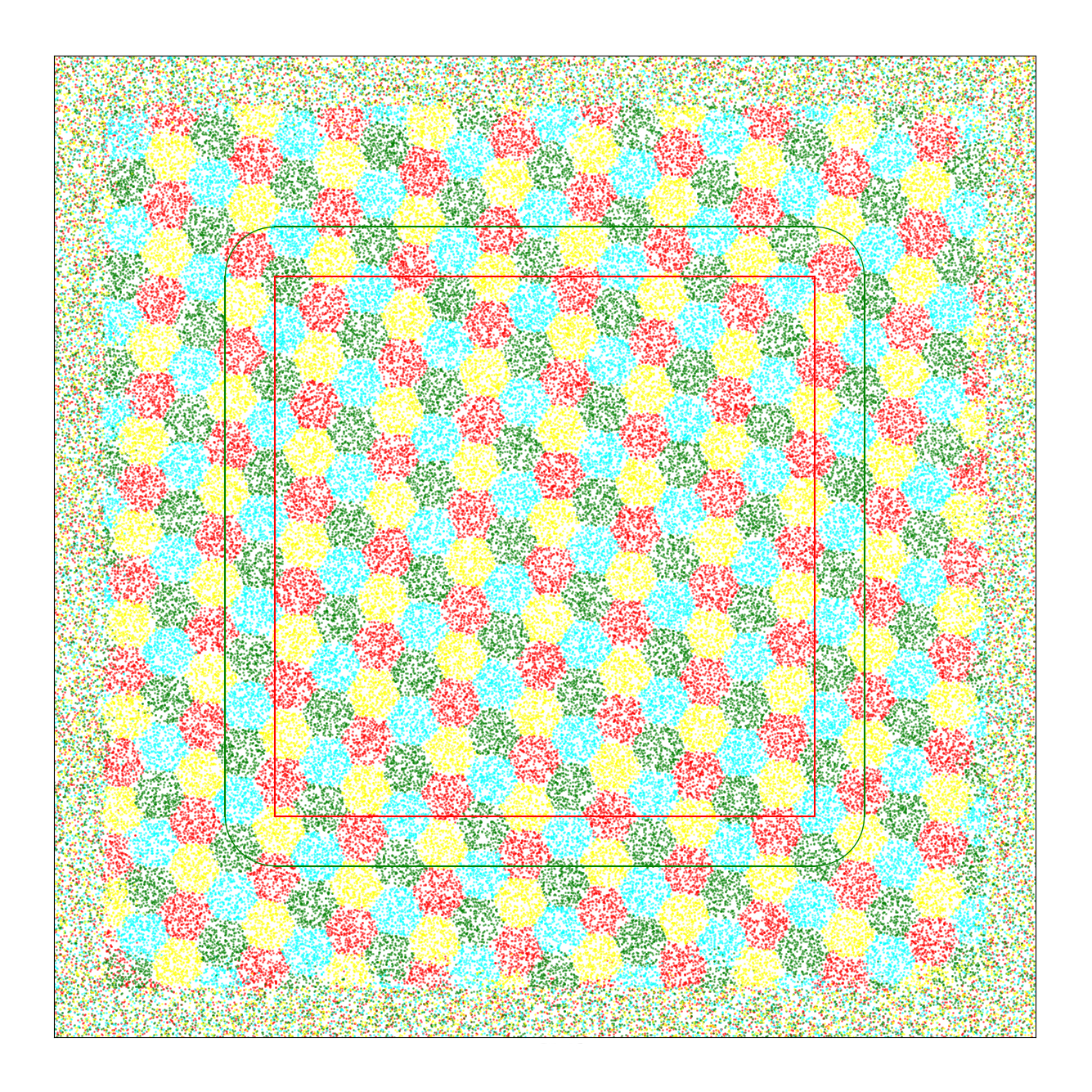}
			\caption{}
				\label{fig:q4}
		\end{subfigure}
		\hfill
		\begin{subfigure}[b]{0.45\textwidth}
			\centering
			\includegraphics[width=\textwidth]{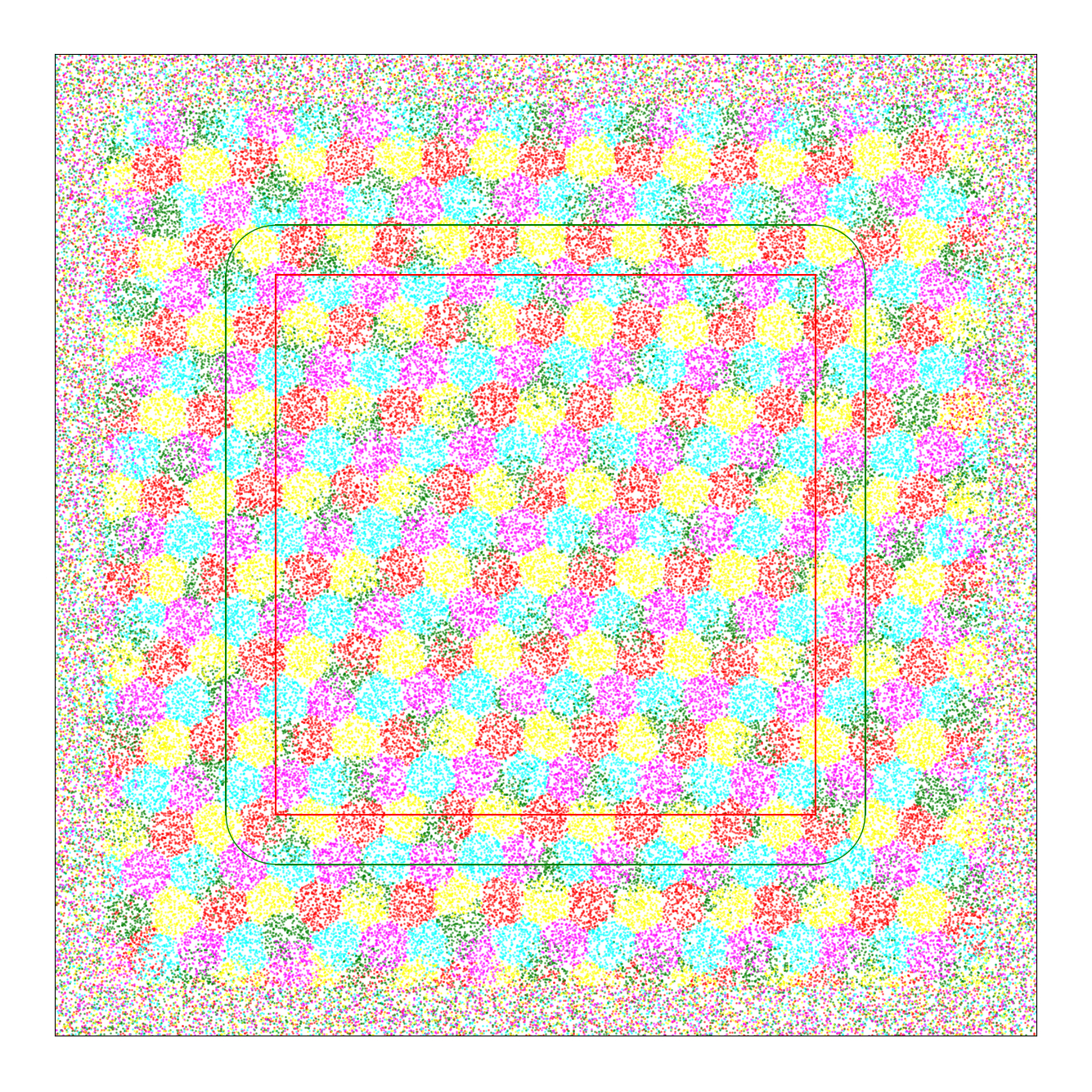}
			\caption{}
			\label{fig:q5}
		\end{subfigure}
		\caption{{\small{Typical vacuum configurations. \textbf{(a)}: $q$ = 4; \textbf{(b)}: $q$ = 5}}}
	\label{fig:q45}
\end{figure}
Let us begin with $q = 2$ colors.
Typical minimized configurations are represented by alternate stripes pattern and shown in the figure \ref{fig:q2}.
The energy of such configuration is about 65\% of the initial configuration energy, corresponding to random coloring.

For $q = 3$ the resulting pattern is hexagonal and consists of regular shapes (figure \ref{fig:q3}).
The minimal energy is about 31\% of the starting configuration.
This number refers to a regular almost pure pattern which emerges in 70\% of minimizations.
It is interesting that the resulting tiling is not an ideal hexagonal lattice.
If one constructs such a lattice with optimal length of hexagon's edge $l = 0.64$ then the energy of it would be higher than energy of minimized configuration, with the ratio of energies of minimized configuration and regular lattice configuration is about 0.86.
Some nontrivial mixing at the borders of the color clusters leads to decreasing of energy comparing to ideal regular hexagonal configuration.

For the case of $q = 4$ the vacuum energy is about 3\% of the initial configuration energy, but this energy is still far from zero value.
The minimized configuration has a regular hexagonal pattern as for $q = 3$ (figure \ref{fig:q4}).
The nearly perfect coloring observed in 40\% of configurations, the other part has some irregularities.
The truly perfect coloring, as well as for $q=3$ does not give a minimum of energy.
Considering optimal $l=0.56$ we get a ratio about 0.7 that is even less than the corresponding ratio for three colors.
It is remarkable how unnoticeable to naked eye changes on the edges of hexagons make such a big effect.
Summarizing results for small $q$'s in the context of EHN problem we would not expect to see zero energy for $q \leqslant 4$.
In the next section we discuss the $q \geqslant 5$ case.

\subsection{$q \geqslant 5$}

\begin{figure}
	\centering
		\begin{subfigure}[b]{0.45\textwidth}
			\centering
			\includegraphics[width=\textwidth]{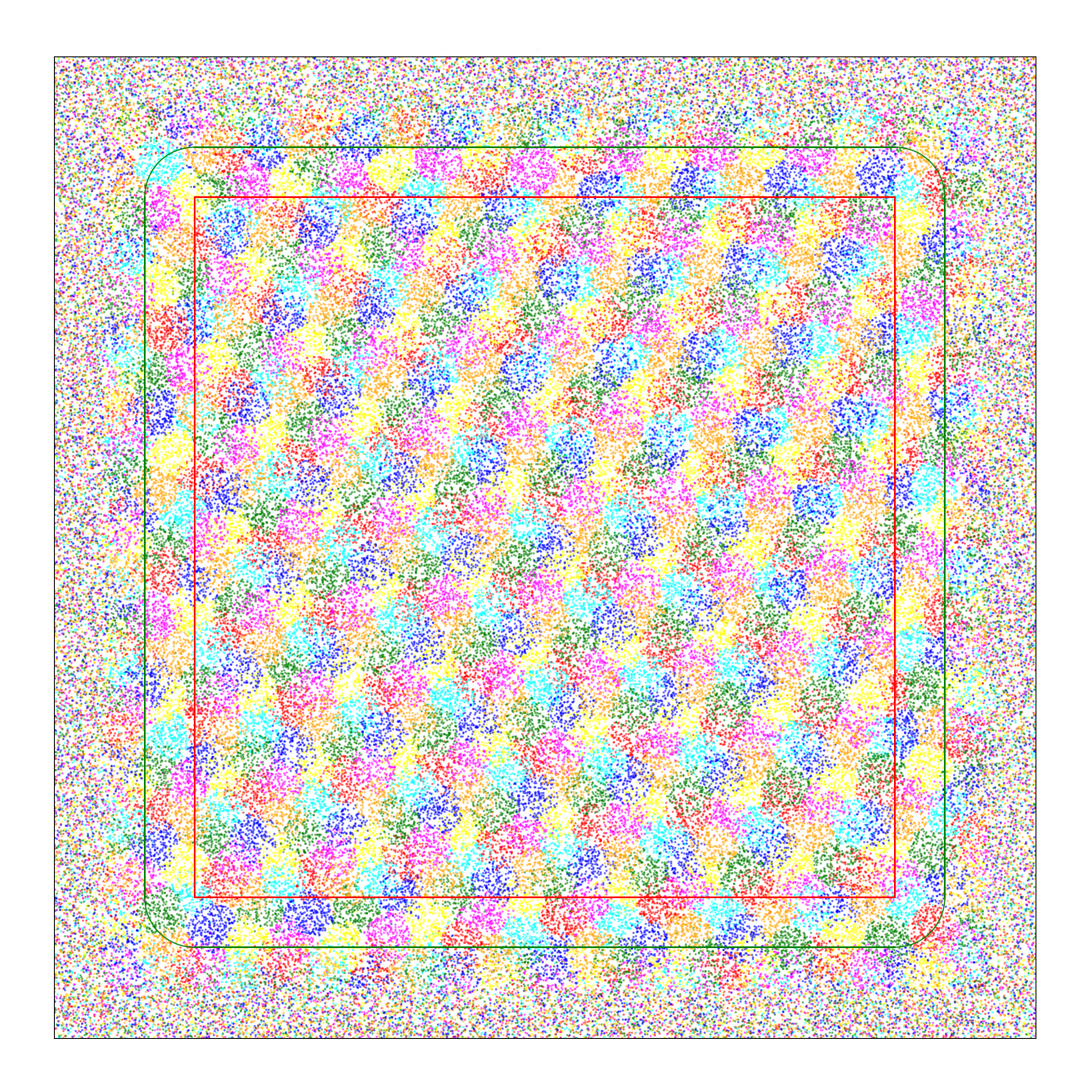}
			\caption{}
				\label{fig:q7}
		\end{subfigure}
		\hfill
		\begin{subfigure}[b]{0.45\textwidth}
			\centering
			\includegraphics[width=\textwidth]{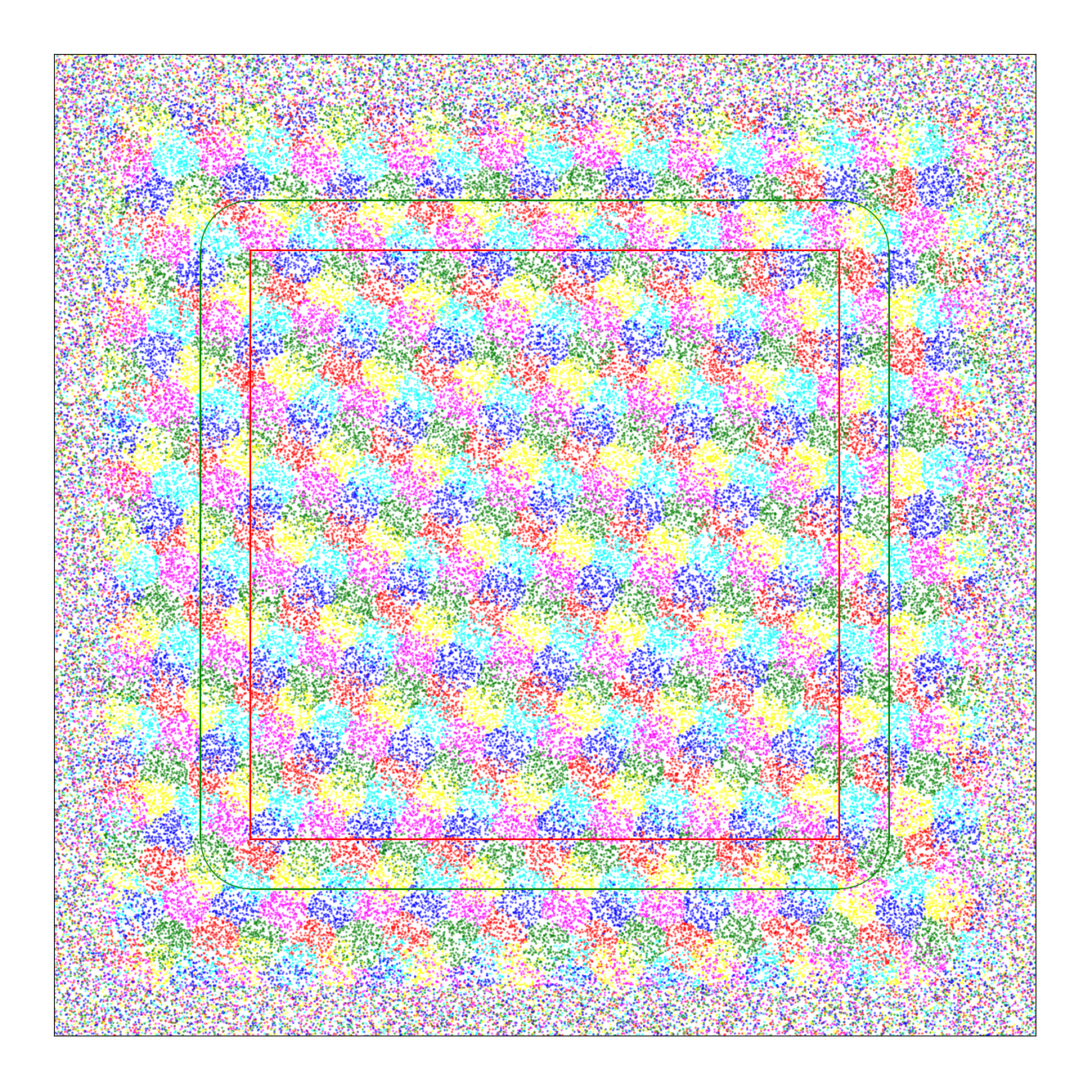}
			\caption{}
			\label{fig:q6}
		\end{subfigure}
		\caption{{\small{Examples of vacuum configurations with $E = 0$. \textbf{(a)}: $q$ = 7; \textbf{(b)}: $q$ = 6.}}}
	\label{fig:q67}
\end{figure}

We will go in reversed order and start with $q=7$ for the reason that it is an upper boundary of the solution to EHN problem and therefore for the discrete version of this we should get zero energy of vacuum configurations.
It is a good test of our model and minimization algorithm.
The tiling in the figure \ref{fig:7colors} allows pretty wide range of changing the length $l$ of the hexagon's edge, $E=0$ for $l \in (\frac{1}{\sqrt7},\frac12)$.
To put on test the simulated annealing algorithm we minimized initial configuration with randomly colored particles and observed zero vacuum energy on 97.5\% of configurations.
The example shown in the figure \ref{fig:q7}.
When we increased the number of neighbors $\langle n \rangle$ by factor 2 (correspondingly increasing $N$), we got zero energy for standard offset on 41\% configurations.
We have found no vacuum configuration with the ideal hexagonal tiling but nevertheless the regular hexagonal pattern is observed.
The result is a juxtaposition of regular lattices of clusters for each color.

\begin{figure}
	\centering
		\begin{subfigure}[b]{0.49\textwidth}
			\centering
			\includegraphics[width=\textwidth]{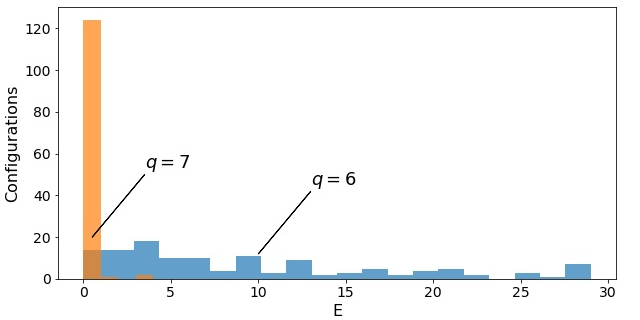}
			\caption{}
				\label{fig:hist67}
		\end{subfigure}
		\hfill
		\begin{subfigure}[b]{0.49\textwidth}
			\centering
			\includegraphics[width=\textwidth]{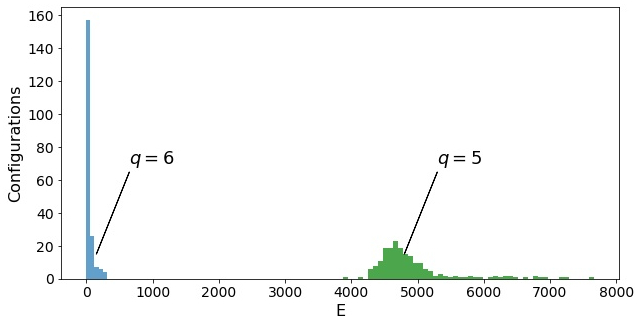}
			\caption{}
			\label{fig:hist56}
		\end{subfigure}
		\caption{{\small{Comparison of distributions of energies for \textbf{(a)}: six and seven colors, 127 the lowest energy configurations; \textbf{(b)}: five and six colors, 200 configurations.}}}
	\label{fig:hists}
\end{figure}

For $q = 6$ the minimized pattern is the same. Configurations with zero vacuum energy arise in about 4\% of cases.
The vast amount of minimizations end up at the vicinity of zero.
The comparison histogram for six and seven colors is shown in the figure \ref{fig:hist67}.
The example of configuration with zero energy is given in the figure \ref{fig:q6}.
It is obvious that if one would construct a regular hexagonal tiling drawing analogy with $q=7$ (figure \ref{fig:7colors}) then the energy of the resulting configuration would not be zero.
What is surprising is how far actually it is from the energy of minimized configurations.
The energy of optimal hexagonal configuration with the edge of hexagon $l = 0.55$ is $E = 15\,740$ in contrast to around zero values for resulting vacuum configurations.

Finally we  move to the description of minimized configurations for $q = 5$.
The typical pattern is shown in the figure \ref{fig:q5}.
Like for $q = 4$ it also has hexagonal tiling but in this case only four colors form the pattern.
The fifth color is displaced.
Corresponding clusters of particles have pretty random shape and number of particles as well as the role of displaced color could go from one color to another in different parts of the area.
Thus we observe the interesting conflict between color and geometric symmetries, and the former got broken as a result of this conflict.
The energy of minimized configuration comprises only about 1\% of energy of starting configurations.
Nevertheless there is a huge gap between distributions of the energies for five and six colors (figure \ref{fig:hist56}).
So based on our numerical results we can conjecture that there is no zero energy vacuum for five colors.
We will discuss this issue below.

%===============improvement=====================
\subsection{The stability of the model}
\label{paramchoice}

\begin{figure}
	\centering
		\begin{subfigure}[b]{0.49\textwidth}
			\centering
			\includegraphics[width=\textwidth]{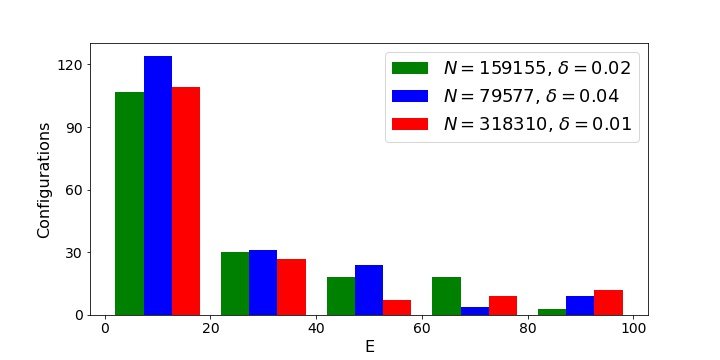}
			\caption{}
				\label{fig:stab6}
		\end{subfigure}
		\hfill
		\begin{subfigure}[b]{0.49\textwidth}
			\centering
			\includegraphics[width=\textwidth]{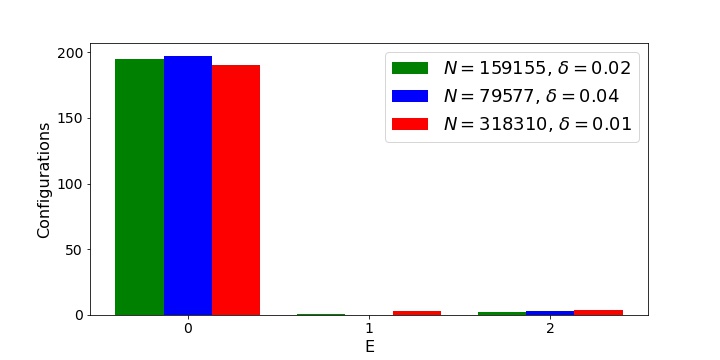}
			\caption{}
			\label{fig:stab7}
		\end{subfigure}
		\caption{{\small{Distributions of minimized energy for different $(N,\delta)$ pairs. \textbf{(a)}: $q=6$; \textbf{(b)}: $q=7$}}}
	\label{fig:stab67}
\end{figure}

As it was said in the section \ref{subsec:1}, we fixed the average number of neighbors $\langle n \rangle = 50$.
Whereas the study of how the variation of $\langle n \rangle$ influences the behavior of the model is not the subject of this article, we will briefly discuss this important parameter.
If one sets $\langle n \rangle$ too small then clusterization does not occur.
From this perspective the choice of such $\langle n \rangle$ that leads to clusterization seems reasonable.
There is no reason to expect clusterization to break down with the increase of the number of neighbors.
The energy minima achieved by simulated annealing algorithm can change but this topic is behind the scope of the article.
The average number of neighbors is controlled by the number of particles $N$ and the width of the interaction ring $\delta$.
By fixing $\langle n \rangle$ we get $N$ and $\delta$ inversely proportional to each other.
The important question is whether the model is stable to variation of these parameters.
In the figure \ref{fig:stab67} three combinations of $N$ and $\delta$ for a fixed number of neighbors are shown.
The energy distributions are similar -- there are some differences but the general pattern is the same which indicates the robustness of the model for a particular number of neighbors.
%===============improvement=====================

\subsection{Breaking of color symmetry}

\begin{figure}
	\centering
		\begin{subfigure}[b]{0.49\textwidth}
			\centering
			\includegraphics[width=\textwidth]{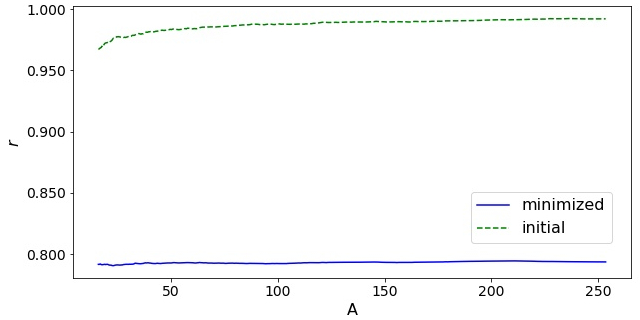}
			\caption{}
				\label{fig:break5}
		\end{subfigure}
		\hfill
		\begin{subfigure}[b]{0.49\textwidth}
			\centering
			\includegraphics[width=\textwidth]{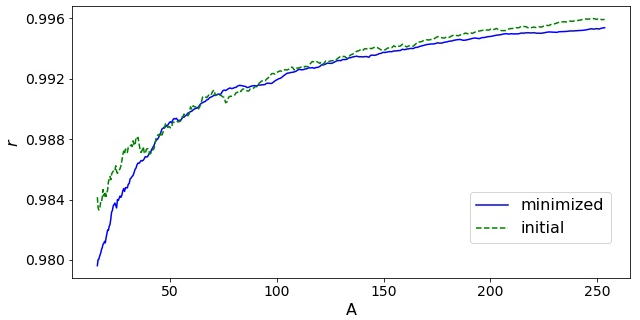}
			\caption{}
			\label{fig:break3}
		\end{subfigure}
		\caption{{\small{The ratio of the number of particles with least represented colors to the total number of particles. \textbf{(a)}: five colors; \textbf{(b)}: three colors.}}}
	\label{fig:break35}
\end{figure}

\begin{figure}
	\centering
	\includegraphics[width=0.65\linewidth]{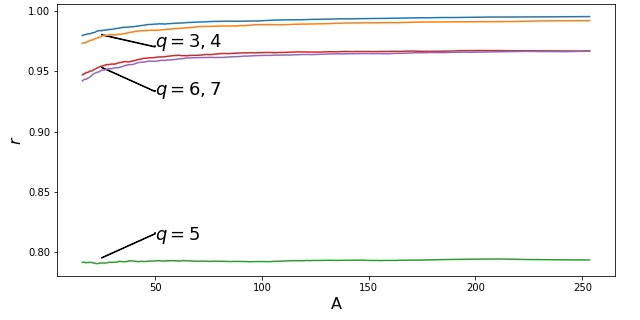}
	\caption{{\small{The ratio of number of particles with least represented colors to the total number of particles for all colors. The three levels are observed. The top two curves correspond to three and four colors, the intermediate ones to six and seven colors whereas the bottom curve represents five colors. }}}
	\label{fig:break_gen}
\end{figure}

The observed color symmetry breaking for $q=5$ is a consequence of the well known fact that no fifth order crystallographic symmetry does exist.
To get quantitative representation of this effect we check  the ratio of particles with least represented color $N_{\mathrm{minc}}(A)$ to whole number of particles $N(A)$ inside particular region.
The ratio scaled by $q$ depends on the area, as seen in the figure \ref{fig:break5}.
The curves for initial random coloring and minimized configurations are quite different.
What interesting is that even for relatively big areas the second curve does not approach the first.
So the effect is very profound and has rather global than local significance.
For comparison one can look at the same graph for $q = 3$ (figure \ref{fig:break3}).
In this case these curves go along together.
The general picture is shown in the figure \ref{fig:break_gen}.
From this we see that color symmetry breaking is not something that is intrinsic only to $q=5$.
Surprisingly it also shows up for six and seven colors (even for $q=4$) but to a far less extent.

\section{Discussion}

We discussed the non-local Potts model in this paper.
We have found peculiar patterns in the ground states, strongly dependent on the number of "colors" $q$ - degrees of freedom of the Potts spins. We believe that such cluster patterns could naturally emerge in any networks with typical link size much larger than the average distance between sites. An interesting example of such a system is given in \cite{article-U} where skin color patterns of lizards are studied. For more examples of this kind see \cite{Anderson}.

There are two most remarkable qualitative observations coming from our numerical analysis.
First, we found color symmetry breaking for $q=5$ corresponding to the absence of degree five symmetry group of the plane. The geometric symmetry outperforms the color symmetry - the system minimizes its energy by quasi-regular pattern with unequal share of one color with respect to four other ones.
Of course, each color can be least represented, the actual choice depends on an initial configuration.
This is to be compared with the Figure 1 from \cite{dum}, where one color dominance was reported in the critical planar Potts model for $q>4$.
Another observation comes from the relation of our model to the well known ENH problem of graph coloring in combinatorial topology.
It was shown analytically \cite{Grey} earlier that four colors are not enough for proper coloring of the plane and our numerical results fully support this conclusion.
Moreover, we have not found a single zero energy configuration for $q=5$.
On the other hand, the zero energy vacuum state known analytically to exist for $q=7$, is clearly seen by our simulations. The situation with $q=6$ needs further refinements and, in ideal case, more systematic work towards the continuum limit (which corresponds to $N\to \infty$, $\delta \to 0$ in our case).
We hope to continue this analysis in future.

\section*{Acknowledgements}
The work was supported by the state assignment of the Ministry of Science and Higher Education of Russia (Project No. 0657-2020-0015). The numerical simulations were performed at the computing cluster Vostok-1 of Far Eastern Federal University.




%\section*{References}
\bibliographystyle{elsarticle-num}      % mathematics and physical sciences
\bibliography{bibliography}   % name your BibTeX data base
%\begin{thebibliography}{00}


%\bibitem{}
%
%\end{thebibliography}
\end{document}